\documentclass[sigconf, nobalancelastpage]{acmart}

\renewcommand\footnotetextcopyrightpermission[1]{}

\usepackage[T1]{fontenc}
\usepackage[utf8]{inputenc}
\usepackage{booktabs}
\usepackage{multicol}
\usepackage{graphicx}
\usepackage{caption}
\usepackage{amsmath} 
\usepackage{balance}
\usepackage{pifont}
\usepackage{multirow}
\usepackage{todonotes}

\def\BibTeX{{\rm B\kern-.05em{\sci\kern-.025emb}\kern-.08emT\kern-.1667em\lower.7ex\hbox{E}\kern-.125emX}}

\begin{document}
\fancyhead{}
\title{Semantic Product Search for Matching Structured Product Catalogs in E-Commerce}

\author{Jason Ingyu Choi}
\affiliation{%
  \institution{Emory University}
}
\email{in.gyu.choi@emory.edu }

\author{Surya Kallumadi}
\affiliation{%
  \institution{Home Depot}
}
\email{surya_kallumadi@homedepot.com}

\author{Bhaskar Mitra}
\affiliation{%
  \institution{Microsoft}
}
\email{bhaskar.mitra@microsoft.com}

\author{Eugene Agichtein}
\affiliation{%
  \institution{Emory University}
}
\email{eugene.agichtein@emory.edu}

\author{Faizan Javed}
\affiliation{%
  \institution{Home Depot}
}
\email{faizan_javed@homedepot.com}

\renewcommand{\shortauthors}{J. Choi et al.}

\begin{abstract}
Retrieving all semantically relevant products from the product catalog is an important problem in E-commerce. Compared to web documents, product catalogs are more structured and sparse due to multi-instance fields that encode heterogeneous aspects of products (e.g. brand name and product dimensions). In this paper, we propose a new semantic product search algorithm that learns to represent and aggregate multi-instance fields into a document representation using state of the art transformers as encoders. Our experiments investigate two aspects of the proposed approach: (1) effectiveness of field representations and structured matching; (2) effectiveness of adding lexical features to semantic search. After training our models using user click logs from a well-known E-commerce platform, we show that our results provide useful insights for improving product search. Lastly, we present a detailed error analysis to show which types of queries benefited the most by fielded representations and structured matching.
\end{abstract}

\maketitle

\section{Introduction}
E-commerce platforms and web search engines share similar goals, which is to display or recommend relevant items (e.g. web documents or products) for a search \citep{amazon_semantic}. However, because product catalogs are more heterogeneous and structured compared to web documents, encoding such documents presents new challenges. For instance, products can have several structured and unstructured fields such as title, description and metadata. Each field can be further divided into multiple instances (e.g. long description, short description, dimensions, units), which vary significantly across product domains \citep{nrmf}. To be successful, models first need to understand the semantics of each field, and utilize fielded representations to perform structured matching between query and document.

Before displaying ranked products to users, typical e-commerce platforms undergo two phases: (1) candidate generation; (2) candidate re-ranking \citep{amazon_semantic}. This work focuses on the candidate generation phase. Our goal is to retrieve all relevant products, which is equivalent to maximizing the recall. To accomplish this, we propose a structured matching module (SMM) that leverages multiple fielded representations to learn a structured matching function. Our method has two advantages. First, SMM utilizes a bottom-up approach to encode instances in each field and transforms these partial representations into an overall document vector. This is more effective when compared to encoding long, heterogeneous documents in one shot. Second, because the encoder is trained to learn query and document representations separately, production systems can generate candidates faster by pre-computing product embeddings.

For evaluation, we trained and validated our model using two data sources in the home-improvement domain: (1) internal user click logs; (2) product search relevance (PSR) dataset. The click logs dataset is sub-sampled from our private click logs. For reproducibility, we selected PSR dataset, which is in the same e-commerce domain but has human-annotated relevance labels. After evaluating our model on these two datasets, we show that incorporating SMM after pre-trained transformer improves the overall matching performance. Our contributions are:

\begin{itemize}
    \item A new structured matching module (SMM) that extends Siamese transformer structures by incorporating fielded representations and lexical signals.
    \item A large-scale empirical evaluation, demonstrating promising performance of SMM for semantic product search in e-commerce.
\end{itemize}
Next, we review related work to place our contributions in context.

\begin{figure*}[ht]
\includegraphics[height=185pt]{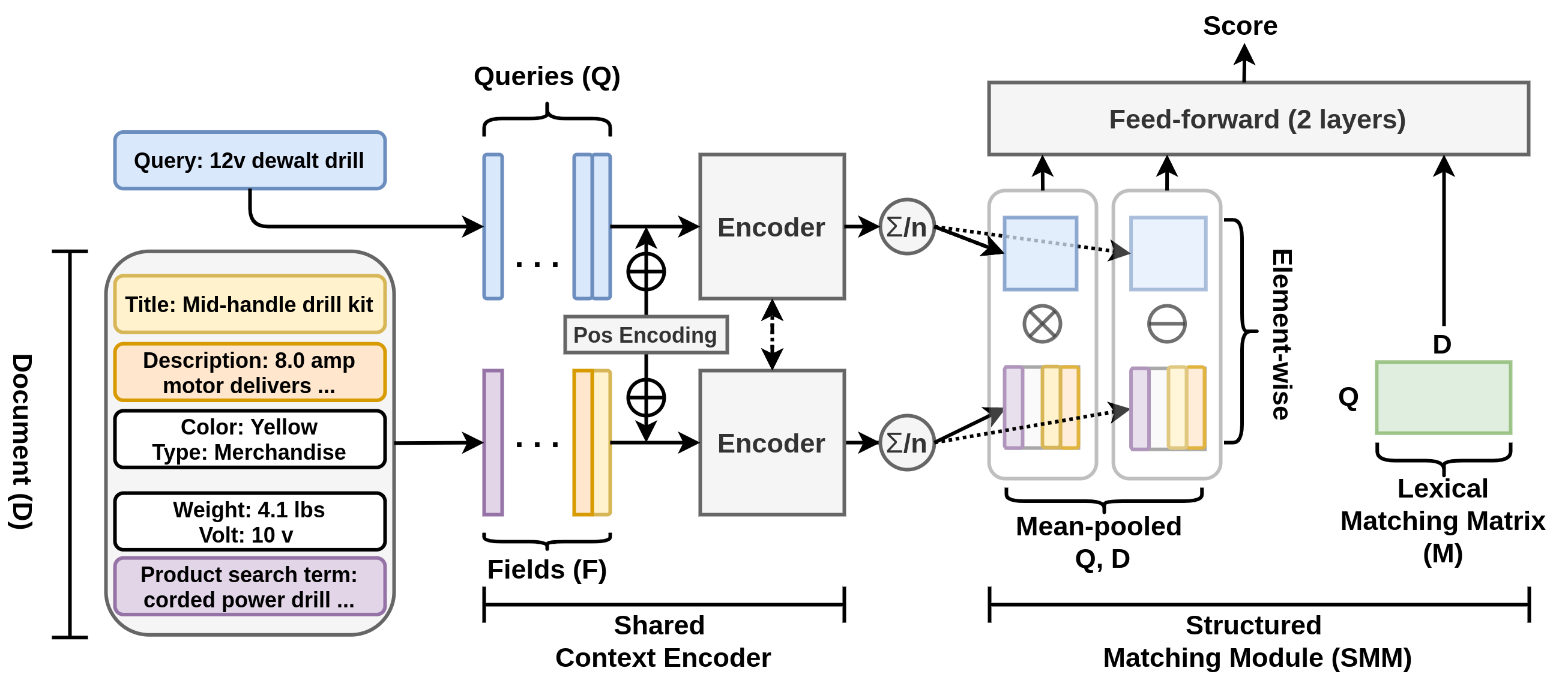}
\captionof{figure}{Overall architecture of our structured semantic matching model.}
\label{model}
\end{figure*}

\section{Related Work}
Traditional ranking approaches combined probabilistic signals and lexical features such as term frequency and document length to rank documents according to their relevance \citep{bm25, bm25f}. However, the main challenges of lexical ranking models are on their coverage since these models cannot make any semantic connection. Distributional semantics \citep{mikolov} was introduced to solve this challenge by training a language model using latent features to obtain a vector representation for each word. 

The earliest models were representation-based models that encode two texts into a fixed-size vector, and computes the similarities by taking cosine or dot product between two vectors \citep{dssm, arcI}. However, these methods were limited because the network did not consider any interaction between the two inputs. To address this problem, interaction-based models were introduced to capture more complex patterns between query and document matrix \citep{knrm}. To combine interactions with lexical signals, hybrid models were proposed and validated the effectiveness of lexical matching when combined with semantic matching \citep{duet, nvsm}. Several approaches evaluated the potentials of incorporating structured information to a document ranking task \citep{nrmf, bm25f}. 

Recently proposed text matching approaches adopt transformer-based encoders to benefit from their rich semantics. However, since transformers use cross-encoder structures, there have been several attempts to decouple the two inputs to learn semantically meaningful vectors for scalable comparisons \citep{siamese_acl, sentence_bert, stn2_trec}. In our work, we maintain the Siamese structure to benefit from these advantages.

\section{Proposed Approach}
In this section, we formally explain the problem and present an overview of our proposed architecture, described in Figure \ref{model}. Our model is composed of two sub-modules: (1) context encoders; (2) structured matching modules. 

\paragraph{\bf Problem and product catalog description}
\label{pd}
Given a user query $Q$ and product $D$, we define product fields $F = \{F_1 ... F_n\}$ and field instances $I = \{F_{1i} ... F_{nk}\}$. All products share the same fields as described in Table \ref{fields}. Instances are attributes of each field that represent certain aspects of a product. For all products, the number of fields $n$ is fixed to 7 while the number of instances $k$ varies between fields and products. The last field titled \underline{Product search terms} is collected from previously issued queries that resulted in a click on this product. We sampled these queries from the training data, and selected the top 10 unique queries based on frequency. In addition, we highlight that \underline{Title} field does not have other instances except the product title. 

\vspace{-3mm}
\begin{table}[ht]
    \centering
    \begin{tabular}{l|l}
    \toprule
    Fields  & Instances  \\
    \toprule
    Title & Product title \\
    Description & Types of descriptions (short, long)  \\
    Product category & Category  \\
    Metadata & Additional descriptions (color, texture)  \\
    Brand & Brand information (brand name) \\
    Numeric & Numeric values (height, width)\\
    Product search terms & Top-10 queries from click logs \\
    \bottomrule
    \end{tabular}
    \caption{Seven fields selected for representing products.}
    \label{fields}
\end{table}
\vspace{-6mm}

For this study, we represented each field $F_i$ as a concatenated sequence of tokens from its instance group. Based on the input pairs $(Q, D=[F_1 ... F_n])$, the goal is to train a function $f$ that maps this input into a probability score $s$, indicating the likelihood of relevance.

\vspace{-5mm}
\begin{gather}
    f(Q, D) \in [0, 1]
\end{gather}

\paragraph{\bf Context encoders}
When encoding query and fields, we used a pre-trained DistilBERT \citep{distilbert}, which is a distilled version of BERT that retains 97\% of the original performance. We chose this model over the original BERT because DistilBERT required less GPU memory and has faster convergence. The weights of transformer blocks are initialized from a pre-trained model, and are tuned using our data. Our encoder used 6 hidden layers with 12 attention heads. 

To obtain the sentence embeddings, we used the mean pooling strategy that takes the mean of hidden states for each sequence position from the final transformer block. This method has shown to be effective compared to directly using the vector for [CLS] token or max-pooling, and is similar to extracting bag-of-words representations where words have interacted with others through multiple self-attention layers \citep{siamese_acl, sentence_bert}. These pooled vectors are stacked to form query and document matrices $\hat{Q}$ and $\hat{D}$.

\paragraph{\bf Structured matching modules (SMM)}
The goal of SMM is to extract matching signals from $\hat{Q}$ and $\hat{D}$. Based on previous literature, we applied element-wise multiplication and subtraction to these matrices to generate features \citep{nrmf, sentence_bert}. We emphasize that because $\hat{D}$ is a stack of field vectors, element-wise operations allow different field vectors to interact with query vectors. This is equivalent to performing pairwise comparisons between query and field in a latent space. In addition, we generated a binary matrix $M$ from the query and document tokens to encode lexical matches. These three outputs are concatenated as following:

\vspace{-3mm}
\begin{gather}
    [|\hat{Q}-\hat{D}|; \, \hat{Q} \circ \hat{D}; \, M]
\end{gather}
\vspace{-3mm}

These outputs are then fed to two layers of non-linear transformations. We used ReLU activation and dropout between these layers. Binary cross-entropy was minimized to predict relevant (1) and non-relevant (0) labels using Adam with 1e-4 learning rate and 16 batch size. Learning rate was warmed-up over the first 10\% of our training data, and linearly decayed with 0.01 decay rate. 0.1 dropout probability was used on all transformer layers to improve regularization. We used 0.5 dropout for our feed-forward layers.

\section{Experimental Setup}
In this section, we present an overview of our data collection process, followed by statistics of our training, validation and two test datasets. The labels of the first test dataset were generated from clicks while the second test dataset was manually annotated from three human workers.

\paragraph{\bf Training dataset from internal click logs}
For this study, we used the subset of click logs from a popular E-commerce platform. For each query, we collected the top 100 (product, clicks) pairs, which are ranked by a production system. We define each entry in our dataset as (query, product, clicks) triple. In total, there are 11,650,964 entries, 1,675,630 unique queries and 3,372,715 unique products. Since our goal is to retrieve all relevant products and not necessarily rank them, we defined a click threshold $r$ to distinguish positive and negative pairs. After manual evaluation, $r=5$ was used to convert click values into a binary label. We filtered out queries that do not contain any relevant product, has a length smaller than 3 characters, or contain numeric values only. For products, we filtered out items that contain too few attributes or do not contain important attributes such as \underline{Title} and \underline{Description}. Lastly, we reserved 5,000 unique queries for a validation set and another 5,000 unique queries for a test set.

\begin{table}[ht]
      \centering
          \begin{tabular}{l|l|l|l}
           \toprule
             & \textbf{Training} & \textbf{Validation} & \textbf{Test} \\
            \bottomrule
            \textbf{Entry} & 11,650,964 & 227,276 & 219,728  \\
            \textbf{Unique query} & 258,666 & 5,000  & 5,000 \\
            \textbf{Relevant} & 51.8\% & 50.7\%  & 52.2\% \\
            \textbf{Not relevant} & 48.2\% & 49.3\%  & 47.8\%  \\
            \bottomrule
            \textbf{Unique products} & \multicolumn{3}{c}{384,506} \\
          \bottomrule
         \end{tabular}
      \caption{Click logs training, validation and test statistics.}
      \label{data_stats1}
\end{table} 
\vspace{-8mm}

\paragraph{\bf Human-annotated test set from Kaggle}
In addition to the click logs dataset, we used a publicly available E-commerce dataset titled Product Search Relevance (PSR) dataset, which was released in 2016 as a Kaggle competition by an e-commerce site in home-improvement domain\footnote{\url{https://www.kaggle.com/c/home-depot-product-search-relevance}}. The goal is to perform a more robust evaluation since labels from the first dataset are heuristically generated from clicks. Instead, the labels of PSR dataset is obtained from three human workers where 1 indicates irrelevant, 2 as partially relevant and 3 as perfect match. Each (Q, D) pair was given to at least three workers, and the final scores were averaged. We observed that this dataset lacks negative samples since 83.9\% of the pairs are labeled at least partially relevant (>=2). After rounding off the decimal values, we reduced the labels into three discrete labels $\in [1, 2, 3]$. $r=2.5$ was used to convert labels into binary labels. 

In addition to these ground-truth labels, we recruited one domain expert and asked to annotate 1,000 randomly sampled queries into 6 classes of \underline{Brand/Collection}, \underline{Color/Finish}, \underline{Unit}, \underline{Material}, \underline{Model}, and \underline{Typo} for error analysis. Each class represents whether the query terms contain important keywords that identify specific classes. Given the nature of multi-intent queries, it is possible to have multiple classes (e.g. Black Samsung TV) per each sample. Our annotated results show that 17.4\%, 4.7\%, 21.8\%, 5.7\%, 3.7\%, and 11.7\% queries (with duplicates) belong to each class respectively. 


\paragraph{\bf Baseline models and metrics}
We chose baselines models in two groups: (1) lexical baselines; (2) neural baselines. For lexical baselines, we will report the performance of BM25 and BM25F rankers, which are tuned on our validation set. Please note that we are not using top@k retrieved results from lexical models to do re-ranking task, but the scores from these models are directly computed as a final matching score for test samples. To index the documents, we used an open-source indexing software titled Terrier \citep{terrier}, and indexed using 7 pre-defined fields from Section \ref{pd}. Standard pre-processing steps such as removing stopwords and stemming are applied before indexing. 

For neural baselines, we experimented with an interaction-based Arc-II model and a hybrid model Duet \citep{duet, arcI}. These two models are trained in a pairwise setting to minimize rank hinge loss. To evaluate matching performance, we chose NDCG, MAP and MRR with $k\in[1,5]$, since these metrics capture how accurately our model retrieves the correct items and their respective positions \citep{ir_metric}. The positions are ranked by the output score from our model. All of these models including ours are implemented in PyTorch framework \citep{huggingface, matchzoo}, and hyperparemeters are tuned using the validation set.


\begin{table*}[ht]
  \begin{tabular}{l|cc|cc|cc|cc}
    \toprule
    \multirow{2}{*}{\textbf{Query labels}} &
      \multicolumn{2}{c}{\textbf{NDCG@1}} &
      \multicolumn{2}{c}{\textbf{NDCG@5}} &
      \multicolumn{2}{c}{\textbf{MAP}} &
      \multicolumn{2}{c}{\textbf{MRR}} \\
      & {\textbf{DB}} & {\textbf{Ours}} & {\textbf{DB}} & {\textbf{Ours}} & {\textbf{DB}} & {\textbf{Ours}} & {\textbf{DB}} & {\textbf{Ours}} \\
      \midrule
      Brand/Collection & \textbf{0.276} & 0.263 & \textbf{0.418} & 0.404 & \textbf{0.427} & 0.416 & \textbf{0.479} & 0.474 \\
      Color/Finish & \textbf{0.311} & 0.278 & \textbf{0.429} & 0.394 & \textbf{0.443} & 0.412 & \textbf{0.504} & 0.460 \\
      Unit & 0.264 & \textbf{0.302} & 0.374 & \textbf{0.390} & 0.391 & \textbf{0.407} & 0.449 & \textbf{0.482} \\
      Material & 0.247 & \textbf{0.376} & 0.416 & \textbf{0.438} & 0.447 & \textbf{0.470} & 0.484 & \textbf{0.591} \\
      Model & 0.257 & \textbf{0.283} & 0.364 & \textbf{0.383} & \textbf{0.410} & 0.404 & 0.451 & \textbf{0.460} \\
      Typo & 0.312 & \textbf{0.334} & 0.423 & \textbf{0.428} & 0.454 & \textbf{0.462} & 0.521 & \textbf{0.541} \\
      All others & 0.306 & \textbf{0.317} & 0.422 & \textbf{0.433} & 0.442 & \textbf{0.452} & 0.504 & \textbf{0.517} \\

    \bottomrule
  \end{tabular}
\caption{Error analysis of different types of queries for DistilBERT (DB) and our model with FMM.}
\label{error}
\vspace{-7mm}
\end{table*}

\section{Empirical Results and Discussion}
We report the comparison of our method against other baselines, followed by feature ablation and error analysis.

\paragraph{\bf Results, ablation study and insights}

Table \ref{main_results} shows that our model outperformed all lexical and neural baselines, showing the effectiveness of combining transformers and SMM. Compared to duet, our model achieved 2.20\% improved NDCG@5, 0.93\% improved MAP and 4.14\% improved MRR respectively. For non-transformer models, duet outperformed all other baselines, validating the effectiveness of leveraging distributed and lexical representations.

\begin{table}[H]
    \centering
    \begin{tabular}{l|l|l|l|l}
        \toprule
        \textbf{Models} & \textbf{NDCG@1} & \textbf{NDCG@5} & \textbf{MAP} & \textbf{MRR} \\
        \toprule
        \textbf{BM25} & 0.280 & 0.384 & 0.419 & 0.462 \\
        \textbf{BM25F} & 0.287 & 0.384 & 0.421 & 0.466   \\
        \textbf{ArcII} & 0.285 & 0.380 & 0.412 & 0.465   \\
        \textbf{Duet} & 0.301 & 0.408 & 0.428 & 0.482   \\
        \toprule
        \textbf{Ours} & \textbf{0.309*} & \textbf{0.417*} & \textbf{0.432} & \textbf{0.502*}  \\
        \toprule
    \end{tabular}
    \caption{Performance comparison of our proposed model on PSR test dataset. ``*'' indicates statistical significance of improvement based on two-tailed Student's t-test with $p<0.05$, compared to Duet model.}
    \label{main_results}
\end{table}
\vspace{-7mm}

To measure the gains from FMM, we conducted an ablation study by training a pre-trained DistilBERT (DB) without fielded representations. For this model, documents are encoded as one long text, thus removing any structured matching advantage. According to Table \ref{ablation1}, after adding FMM, we observed statistically significant improvements on MAP and MRR on both test set. Interestingly, we noticed the improvements on NDCG@1 was very small. We hypothesize that FMM does not contribute much to obvious cases but more to harder cases with heterogeneous instances. For significance testing, we used two-tailed Student's t-test with $p<0.05$. 

\paragraph{\bf Error Analysis}
To more understand where FMM helps and fails, we conducted an error analysis to see trade-offs in metrics after adding FMM. Table \ref{error} shows queries containing field-specific terms are harder than general queries since retrieval performances on \underline{All others} class are higher than those of other classes. Among fields, matching numerical units are shown to be the most difficult task based on NDCG. This is true because numbers are usually filtered out before training, and without understanding the conversions between different units, it becomes very challenging for models to match query without sufficient text features. After adding FMM, we observed several improvements over various types of queries. There is a decrease in performance for \underline{Brand/Collection} and \underline{Color/Finish} types, but our model performed better on all other labels. Interestingly, our proposed model improved on \underline{Typo} class, showing the benefits of subword lexical matching. Similarly for \underline{Units} field, knowing the occurrences of unit matches benefited the overall matching performance. To conclude, we claim that our proposed FMM modules reduce the general errors by providing extra evidence from query to field relationships.

\begin{table}[ht]
    \centering
    \begin{tabular}{l|c|l|l|l|l}
        \toprule
        \textbf{Models} & \textbf{Test} & \textbf{NDCG@1} & \textbf{NDCG@5} & \textbf{MAP} & \textbf{MRR} \\
        \toprule
        \textbf{DB} & \textbf{PSR} & 0.304 & 0.411 & 0.428 & 0.488   \\
        \textbf{Ours} & \textbf{PSR} & \textbf{0.309} & \textbf{0.417} & \textbf{0.437*} & \textbf{0.502*}  \\
        \toprule
        \textbf{DB} & \textbf{CL} & 0.682 & 0.696 & 0.647 & 0.785   \\
        \textbf{Ours} & \textbf{CL} & \textbf{0.682} & \textbf{0.712*} & \textbf{0.705*} & \textbf{0.807*}  \\
        \bottomrule
        
    \end{tabular}
    \caption{Ablation study of our proposed model against DistilBERT (DB) baseline after removing FMM on both product search relevance (PSR) dataset and click logs (CL) dataset.}
    \label{ablation1}
\end{table}
\vspace{-5mm}

\paragraph{\bf Conclusions}
We proposed a novel and effective method for matching queries to structured product descriptions for e-commerce search and recommendation. We adopt a state of the art transformers in Siamese architecture to avoid jointly encoding query and documents for improved scalability. Multiple fielded representations are encoded to first form document matrix, and matched against query vectors to extract heterogeneous matching signals. After evaluating our method on two e-commerce dataset, we showed promising directions of representing documents into multiple vectors both rough ablation study and error analysis. Overall, our results provide useful insights into the benefits and limitations of the proposed method, which could further benefit improvements to e-commerce matching, search, and recommendation.

\balance
\bibliographystyle{abbrv}
\bibliography{acmart}
\end{document}